# Optical tweezers with optical vortex based on deep learning

Zhe Shen[*] and Ning Liu

*School of Electronic and Optical Engineering, Nanjing University of Science and Technology, Nanjing 210094, China*
*\*shenzhe@njust.edu.cn*

**Abstract:** Optical tweezers (OTs) with structured light expand degrees of freedom of particle manipulation. However, the studies of structured optical tweezers are usually accompanied by complex theoretical models, strict simulation conditions, and uncertain experimental factors, which may bring about high time costs and insufficiently precise results. In this work, we proposed a bidirectional neural network model for the analysis and design of OTs with optical vortices (OVs) as a typical structured light beam. The deep learning network deriving from the convolutional neural network was optimized to fit the optical vortex tweezers model. In analyzing optical forces, the network can achieve over 98% accuracy and improve computational efficiency by more than 20 times. In further analyzing particle trajectories, the network can also achieve over 95.5% accuracy. Meanwhile, in OTs with OV-like beams, our network can still predict particle motion behavior with a high accuracy of up to 96.2%. Our network can inversely design optical vortex tweezers on demand with 95.4% accuracy. In addition, the experimental results in OTs with plasmonic vortex can be analyzed by the proposed model, which can be used to achieve arbitrary optical manipulation. Our work demonstrates that the proposed deep learning network can provide an effective algorithmic platform for the analysis and design of OTs, and is expected to promote the application of OTs in biomedicine.

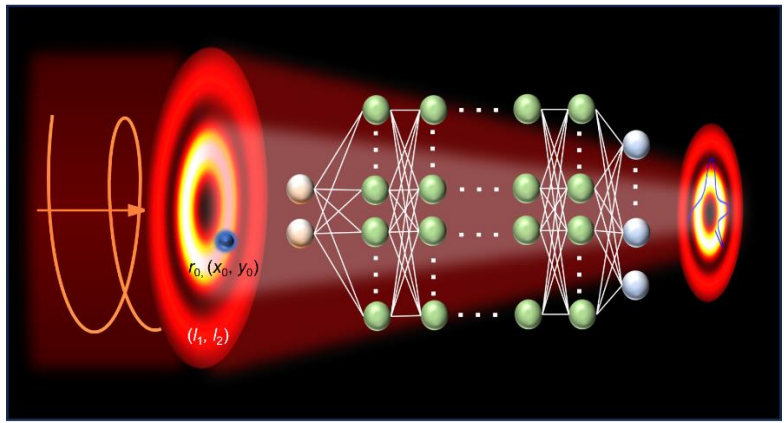

## 1. Introduction

Optical tweezers (OTs) is a technique that enables particle contactless trapping by creating potential wells using a single focused beam, originating from the study of Ashkin and co-workers in 1986 [1]. However, the traditional system with OTs is usually on the basis of a Gaussian beam and requires mechanical control of the light source for achieving complex particle manipulation. In recent years, the rise of structured light has provided more precision and degrees of freedom for complex particle manipulation [2-4]. A variety of extended studies [5-8] have sprung up, such as manipulation of "pushing" [9], "pulling" [10, 11], "rotating" [12, 13], and so on. Optical vortices (OVs) is a typical structured light field carrying helical phase and orbital angular momentum (OAM) [14]. It has been used to stably trap and rotate particles acting as optical spanners [15]. Due to the tunability of topological charge in OVs, it can achieve more flexible manipulation further [16]. Based on these, optical vortex tweezers have been widely applied in optical sorting [7], micromechanics [17], and so on.

For the studies of optical vortex tweezers, in theory, Zhan calculated the optical forces of stably trapped metallic Rayleigh particles in focused radially polarized light [18]. Ng *et al*. estimated the complex motion behavior of dielectric particles in Laguerre-Gaussian (LG) mode OV beams using Mie scattering theory [19]. However, due to the complexity of the theoretical model, obtaining precise analytical solutions needs significant time costs. In simulation, Qin *et al*. studied the effects of particle size, incident wavelength, and polarization state on the optical trap force of metallic particles in tightly focused vector beams using the finite-difference time-domain (FDTD) and T-matrix methods [20]. The method has also been used by Yan *et al*. to study the transfer from OAM to force of a single micrometer silver nanowire in LG beams [21]. However, this method is usually limited by boundary conditions and mesh accuracy, which may lead to some inevitable deviations in the final results. In the experiment, Simpson *et al*. achieved the rotation of micrometer dielectric spheres in LG beams due to the spin-orbit interaction [22]. In our previous work, we observed the steady rotation of a mesoscopic metallic particle trimer in the tightly focused circularly polarized optical vortex [23]. However, these experimental results are inevitably affected by thermodynamics, medium damping, and other such factors. Therefore, obtaining precise solutions between optical vortex tweezers and particle motion is difficult.

Deep learning networks have been proven to have advantages in obtaining precise solutions, and have been greatly developed recently. They are proficient at handling the relationship between large amounts of input and output data to solve high-dimensional data prediction and inverse system design. Compared to traditional methods such as topology optimization, particle swarm optimization, etc. deep learning networks can not only dramatically reduce computational costs, but also improve the accuracy of the results. In the application of artificial intelligence for photonics, deep learning has been applied in predicting the electromagnetic response properties of metamaterials [24], enhancing digital video microscopy [25], predicting the scattering of nanoparticles [26], and estimating microrheological properties [27]. As an important branch in photonics, OTs have been studied with some attempts through deep learning networks [28, 29]. However, the attempts generally suffer from the simplicity of using deep learning networks and the lack of consideration of OTs environment.

In this work, we proposed a bidirectional deep-learning network framework to analyze and design optical vortex tweezers for the first time. The deep learning network consists of a forward network and an inverse network to support the analysis and design of optical vortex tweezers, respectively. Firstly, we analyzed the optical forces of particles in optical vortices through a forward subnetwork. Secondly, we studied the motion trajectories of particles by a forward primary network. Thirdly, we designed the optical vortex tweezers according to pre-defined particle trajectory by using the inverse network. Moreover, we studied particle trajectory in OV-like beams in the form of Archimedean spirals by the forward primary network.

Finally, we analyzed the experimental results in OTs with plasmonic vortex by the proposed network. Our work demonstrates our neural network frameworks can provide an effective algorithm platform for the high-precision analysis and design of OTs systems, and this platform is expected to bring more possibilities for the applications of drug delivery, microfluidics, and so on.

## 2. Method and Theory

Our proposed deep learning framework is a bidirectional neural network. As shown in Fig. 1, it consists of a forward network and an inverse network. The forward network contains a primary network and a subwork. The primary network is shown in the blue solid frame, and it contains a one-layer convolutional layer, two repeated up- and down-sampling blocks, and a fully connected layer to extract data details while increasing data dimensions. In the primary network, blue arrows are used to indicate the direction of forward data flow, and the network input and output are the parameters of optical vortex tweezers and particle trajectory, respectively. The subnetwork is shown in the purple dashed frame, which is a link of the primary network to predict optical forces by inputting different parameters of optical vortex tweezers. The inverse primary network is shown in the red solid frame, and it contains two repeated one-dimensional convolutional residual layers and a fully connected layer to achieve feature extraction while reducing data dimensions. In the inverse network, red arrows are used to indicate the direction of inverse data flow, and the network input and output are particle trajectory and parameters of optical vortex tweezers, respectively.

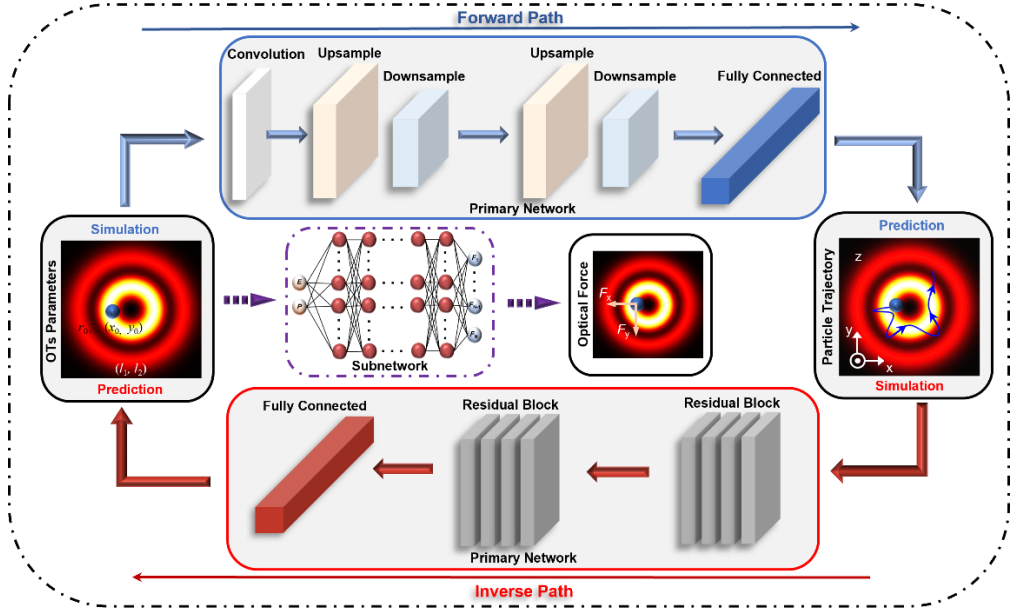

**Fig. 1.** A flowchart of the deep learning network analyzing and designing optical vortex tweezers. Inside the blue solid frame is the forward primary network, inside the purple dashed frame is the forward subnetwork, and inside the red solid frame is the inverse primary network.

During the training of the primary network, the input data dimensions will be continuously expanded, which may result in severe parameter mismatch [30]. It is difficult to solve this problem when using only the fully connected layer. Therefore, we added up- and down-sampling blocks before the fully connected layer. The detailed composition of up- and down-sampling blocks are shown in Figs. 2(a, b), respectively. In the up-sampling block, the input is first mapped as $H_1$ through the deconvolutional (Deconv) layer. Then, $H_1$ is mapped as $H_2$ through the convolutional (Conv) layer. The residual between input and $H_2$ is mapped as $H_3$ through the deconvolutional layer. The final output is obtained by summing $H_1$ and $H_3$. During the flow of data through the up-sampling block, the input data can be continuously integrated and extracted detailed features, while increasing the dimension of the input data reasonably. The output of the up-sampling block will be used as the input of the down-sampling block. The down-sampling block can be obtained by swapping the order of convolutional and inverse convolutional layers in the up-sampling block. The global information can be obtained after the input data flows through the down-sampling processing, and the input data will be reduced in both dimension and size. Through iterating the up- and down-sampling blocks, we can obtain enough information in the input data to reduce the effects of parameter mismatch without changing the data dimensions. Similar models have been used to implement non-linear mapping from low-resolution input to high-resolution output [31]. In addition, in order to further solve the problem of parameter mismatch, we simultaneously utilized the idea of greedy training hierarchy [32]. So, we adopted a two-stage training method in the network, and both segments used the same network architecture.

In the inverse network, we employed a typical convolutional neural network structure, i.e., residual network, which is used to feature extraction of a pre-defined particle trajectory. The detailed composition of the residual network is shown in Fig. 2(c). The input is mapped as $P_1$ through a convolution layer with kernel size $1 \times 1$, and mapped as $P_2$ through two convolutional layers with kernel size $3 \times 3$. The final output is obtained by summing $P_1$ and $P_2$. After continuous residual network processing, we can effectively capture the features in the particle trajectory to improve the accuracy of predicting the parameters of optical vortex tweezers.

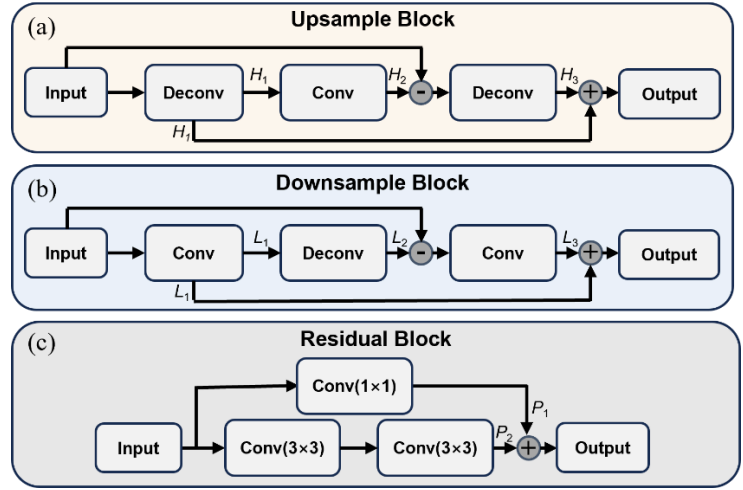

**Fig. 2.** Detailed algorithm diagram of (a) up-sample block, (b) down-sample block, and (c) residual block.

To obtain the initial data for training the deep learning network, we need to theoretically calculate the optical forces and particle trajectories of different optical vortex tweezers. Firstly, we calculated the optical forces using optical tweezers computational toolbox (OTT) based on the Lorentz-Mie model combined with the T-matrix method [33]. Then, based on the optical forces, we calculated the motion trajectories of particles. Here, we ignore the Brownian motion of particles for simplicity. The motion equation of a particle in optical vortex tweezers can be expressed as [19]：

$$F_{\text{light}}(\mathbf{r}) - F_f = m\frac{d^2\mathbf{r}}{dt^2}. \tag{1}$$

where $F_{\text{light}}(\mathbf{r})$ represents the optical force, $m$ is the particle mass, and $\mathbf{r}$ is the position vector that deviates from the equilibrium. where $F_f$ is the resistance of the particle in the ambient medium, which can be calculated through $F_f = \gamma(d\mathbf{r}/dt - v)$. $v$ and $\gamma = 6\pi a\mu$ are the velocity of the medium flow and the ambient damping constant, respectively. Here, $a$ is particle radius and $\mu$ is the dynamic viscosity of the ambient medium.

## 3. Results and Discussions

### 3.1. Prediction of optical forces by forward subnetwork model

Prediction of optical forces using networks is essential for studying optical tweezers based on deep learning, since optical force is a key intermediate variable for obtaining particle trajectory according to Eq. (1). Thus, we used a forward subnetwork to predict optical forces. Before training the subnetwork, we built the optical vortex tweezers model for collecting the data of the design parameters and optical forces, to obtain a database. In model settings, at the case of the input power $p = 100$ mW, the incident light is OV with the wavelength of 1064 nm, tightly focused through an objective lens with numerical aperture NA = 1.2. Spherical particles with refractive index $n = 1.52$ in water were subjected to optical forces in tightly focused OVs. Here, variable parameters of optical vortex tweezers are: the particle radius ($r_0$) is in the interval [100 nm, 550 nm], the particle initial position ($x_0$, $y_0$), where $x_0$ is in the interval [-1 μm, -0.5 μm] and $y_0 = 0$ μm, the type of optical vortex field using LG mode, and these types are distinguished by topological charge ($l_1$, $l_2$).

The subnetwork shown in Fig. 1 is a typical multilayer perceptron (MLP) with four hidden layers, where the number of nodes in each hidden layer is set to 3200, 1600, 800, and 400 respectively. It allows data to flow between the 4-dimensional parameters of optical vortex tweezers and the 300-dimensional optical forces and builds a mapping. Due to the central symmetry of the optical vortex field, the optical force on the particle at any position in the optical vortex field can derived by extrapolating the linear force field distribution around the singularity at the center. Therefore, when we took the distribution of the force field at $y_0 = 0$ as the output data, we can ignore the particle position and reduce the input parameters to 3 dimensions. Based on this, we used training sets in the database to train the subnetwork for achieving convergence conditions, and used testing sets to test. Figs. 3(a, b) show the curve of the optical forces in the x- and y- directions on the particle at position $y_0 = 0$ by simulation and prediction, respectively. It can be observed a high consistency between the computed results of the two approaches, and the accuracy is over 98%. The accuracy percentage was obtained from the difference between the simulated and predicted data. Meanwhile, we quantitatively compared the computational costs of the standard T-matrix method and the subnetwork and found that the computational efficiency of the latter is several tens of times higher than the former. Specifically, the subnetwork can directly reduce the time from 7.13 s to 0.25 s during the calculation of a set of data. This advantage of computational efficiency will be highlighted when acquiring large amounts of data. In short, it is verified that optical forces can be obtained precisely while greatly reducing the time cost by using the proposed subnetwork, which lays a foundation for the acquisition of complex particle trajectories.

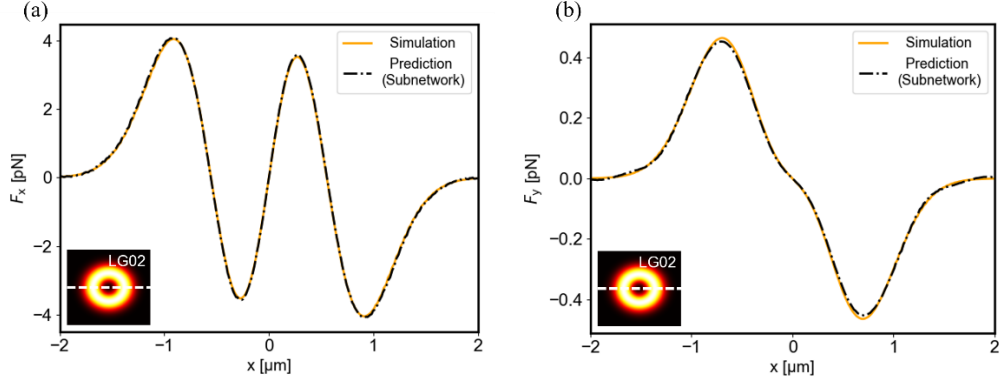

**Fig. 3.** The curves of optical forces along the (a) x-axis and (b) y-axis using simulation and prediction along the middle of the optical field (insert), indicated by the white line.

*3.2. Prediction of particle trajectory by forward primary network model*

After predicting optical forces by the subnetwork, we can continue the analysis of particle trajectory by the forward primary network. Based on the predicted optical force in the subnetwork, we calculated the particle trajectory in the optical vortex tweezers model using Eq. (1) and replaced the optical forces data of the previous database with it to build a new database. The parameters are the same with those in section 3.1. To evaluate the prediction ability of the forward primary network, we used training sets to train the forward primary network and used testing sets to test it. The evaluation results of the forward primary network are shown in Fig. 4. We first predicted the positions along the x- and y- axis over time, respectively, and the positions were used to generate particle trajectory. Here, the data dimension of position along the x-axis is consistent with that along the y-axis, i.e., both are 300 dimensions. Fig. 4(a) shows our arbitrarily selected set of design parameters of optical vortex tweezers with $(l_1, l_2) = (0, 3)$, $r_0 = 100$ nm, and $(x_0, y_0) = (-0.7$ μm, $0$ μm). They were used to evaluate the performance of the forward primary network in this strategy. Fig. 4(b) shows the prediction result of particle trajectory. Since the forward primary network can accurately predict the position along the x- and y-axes over time, as shown in Figs. 4(c, d), the simulated particle trajectory and predicted particle trajectory have high consistency, and the prediction accuracy is over 90%. This result indicates that the forward primary network can effectively establish a mapping between parameters of the optical vortex tweezers and particle trajectory in this strategy.

Further, to simplify the above strategy, we combined the positions along the x- and y-axis as output results of the forward primary network. Due to the dimension expansion from previously 300 dimensions to 600 dimensions, we adopt a two-stage training approach here, the first segment extends the 4 dimensions to 60 dimensions, and the second segment extends the 60 dimensions to 600 dimensions. Figs. 4(e, f) show the prediction results of the particle trajectory by the forward primary network in two sets of arbitrarily selected design parameters of the optical vortex tweezers. The two sets of design parameters are $(l_1, l_2) = (1, 2)$, $r_0 = 100$ nm, $(x_0, y_0) = (-0.5$ μm, $0$ μm) and $(l_1, l_2) = (1, 3)$, $r_0 = 550$ nm, $(x_0, y_0) = (-0.65$ μm, $0$ μm), respectively. We found that the forward primary network can accurately predict the motion position of particles at different moments, and the prediction accuracy exceeds 95.5%. In short, our forward primary network is effective for obtaining particle trajectories by directly predicting two-dimensional variables.

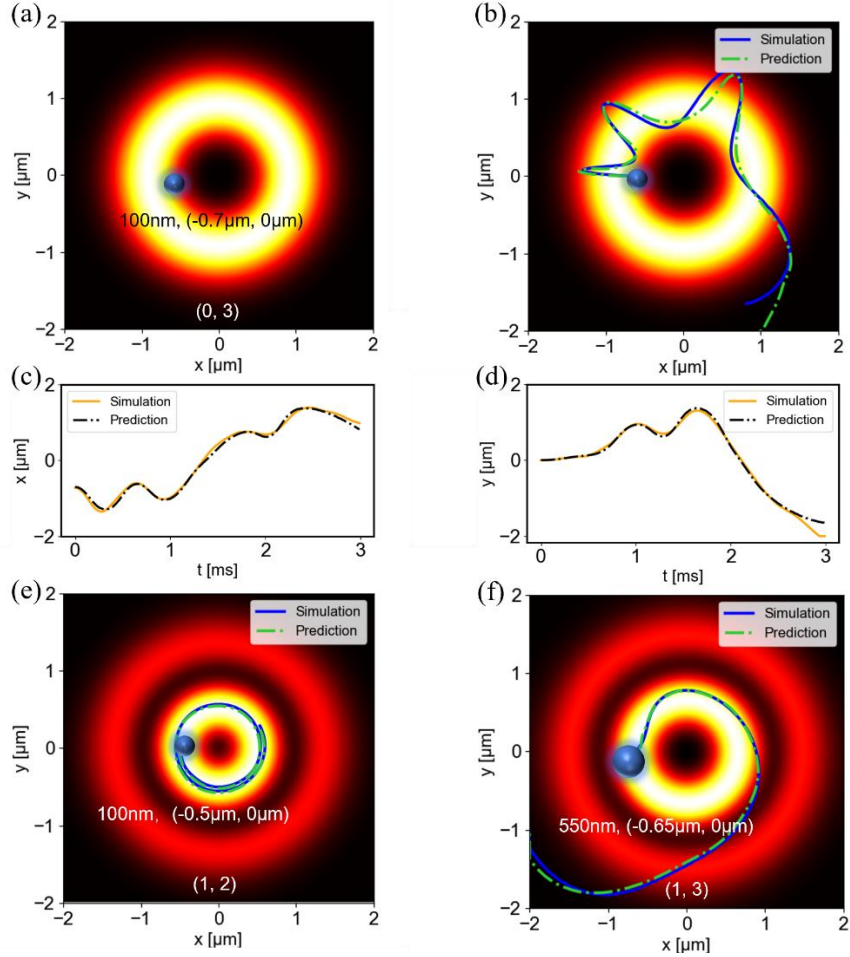

**Fig. 4.** The evaluation results of the forward primary network, predicting particle trajectory. (a) Design parameters of OTs with LG mode OVs. (b) The prediction results of particle trajectory by combining the coordinates in the x-axis and y-axis. The prediction details of coordinates in the (c) x-axis and (d) y-axis over time. (e, f) The comparison results between simulated particle trajectory and predicted trajectory by the forward primary network.

*3.3. Design of optical vortex tweezers by inverse primary network model*

Besides obtaining particle trajectories from the specified optical vortex tweezers, it is of great significance to achieve the particle trajectory on demand, i.e., design the optical vortex tweezers. Therefore, we used the inverse deep learning network shown in Fig. 1 to design the optical vortex tweezers by pre-defined particle trajectory. Figs. 5(a, b) show the comparison of particle trajectory from designed optical vortex tweezers and pre-defined ones. In the test, the arbitrarily selected two sets of target parameters of optical vortex tweezers are $(l_1, l_2) = (0, 1)$, $r_0 = 460$ nm, $(x_0, y_0) = (-0.95$ μm, $0$ μm$)$ and $(l_1, l_2) = (2, 2)$, $r_0 = 258$ nm, $(x_0, y_0) = (-0.1$ μm, $0$ μm$)$, respectively. Since the inverse primary network can accurately predict the parameters of optical vortex tweezers, as shown in Figs. 4(c, d), the two sets of comparison of particle trajectory from designed optical vortex tweezers and pre-defined ones have low errors. In short, our proposed inverse primary model can effectively optimize and design optical vortex tweezers on demand, which is expected to be applied in drug delivery, microfluidics, and so on.

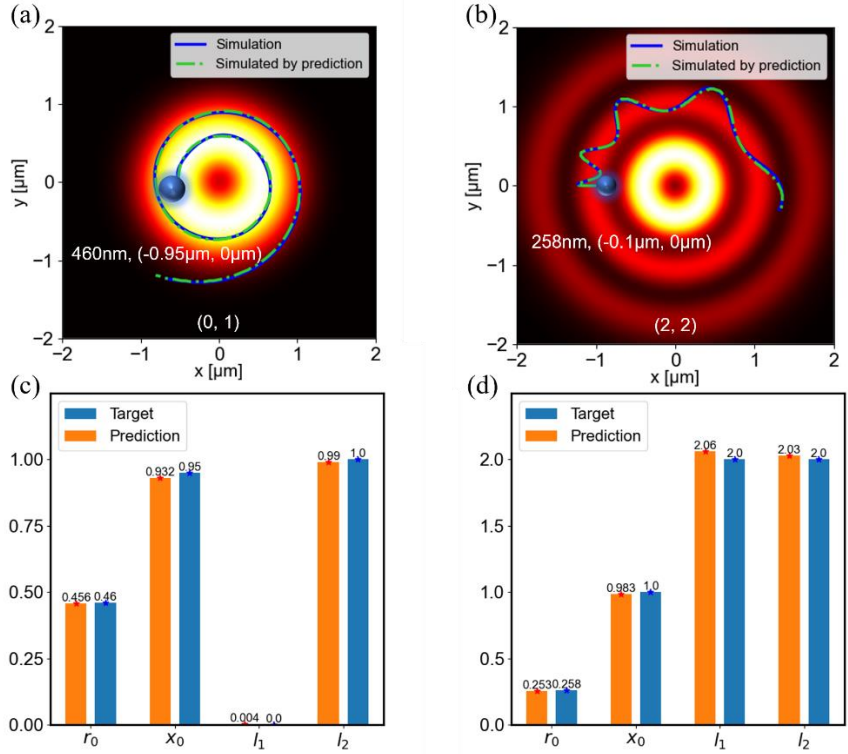

**Fig. 5**. Inverse design of OTs with LG mode OVs using the inverse primary network. (a, b) The comparison results of particle trajectories calculated by simulation between target design parameters and the design parameters predicted by the inverse network. (c, d) The comparison of parameter values in optical vortex tweezers between the target and designed by the inverse network.

*3.4. Prediction of particle trajectories in OV-like beams*

In previous sections, we have analyzed and designed optical vortex tweezers based on the deep learning network. In this section, to explore the applicability of our network, we analyzed particle trajectory in OTs with OV-like beams using our forward primary network. OV-like beams are a class of beams that obtain certain optical field distributions by modifying OVs [34-36], and they have helical phases to drive the particles rotating along the light field distribution. Here, the OV-like beam is the Archimedean spiral beam. Fig. 6(a) shows the Archimedean spiral-beam optical field distribution and variable parameters in the OTs model with the Archimedean spiral beam. The setting range of the variable parameters can be set as follows: the particle radius ($r_0$) is in the interval [100 nm, 500 nm], the particle initial position ($x_0$, $y_0$), where $x_0$ is in the interval [-0.06 μm, 0.06 μm] and $y_0 = 0$ μm, and other parameters are the same as those in section 3.1. The motion trajectory shown in Fig. 6(b) of the particles fits perfectly with the Archimedean spiral beam designed, which is exactly the result of our desired selection. Due to changes in design parameters, we can also obtain imperfect motion trajectories of particles shown in Figs. 6(c, d). More detailed particle parameters are labeled in Figs. 6(c, d). Then, we used the above data to evaluate our network. As shown in Figs. 6(b-d), the trained primary network can make high-precision predictions regardless of whether the particle's trajectory along an Archimedean spiral orbit is perfect or imperfect. This result demonstrates our forward primary can effectively predict particle trajectory in the Archimedean spiral beam. Further, it is foreseeable that our deep learning model may be applicable in other complex structured light.

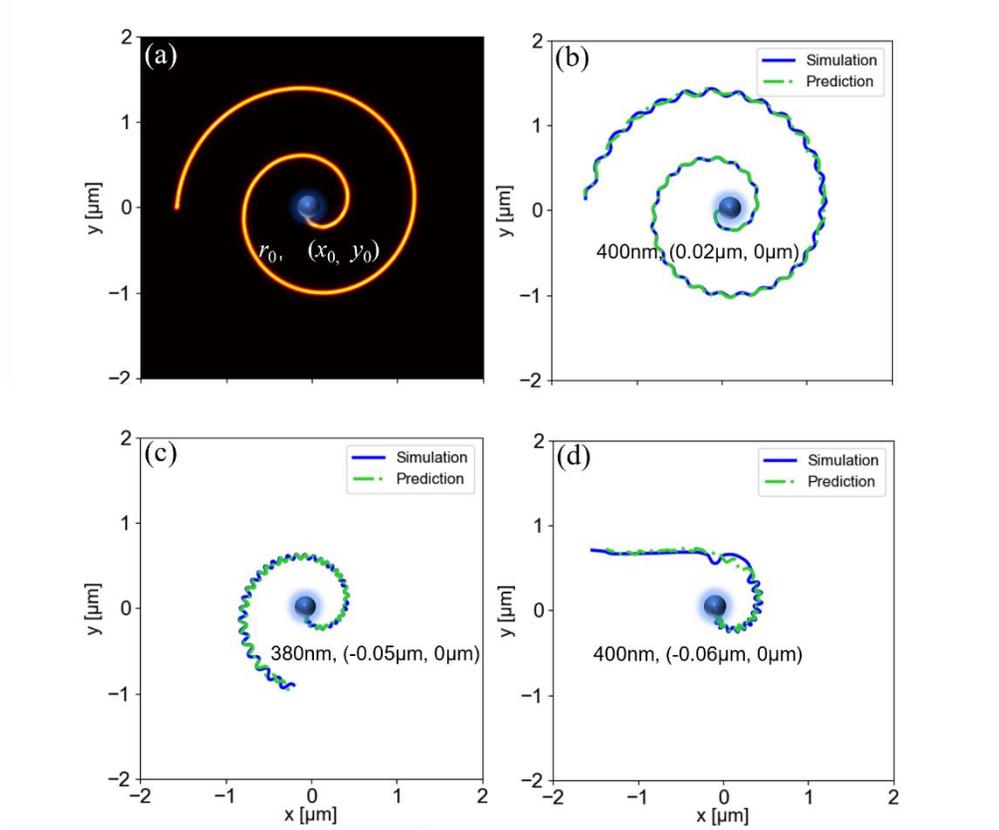

**Fig. 6.** (a) Optical field distribution of the Archimedean spiral beam projected in the focal plane (z = 0). (b-d) The comparison results between simulated particle trajectory and predicted trajectory of the forward primary network.

### 3.5 Analyzing optical tweezers with plasmonic vortex in experiment using deep learning network

To verify that the proposed deep learning network is effective for optical vortex tweezers design in practical application, we analyzed the experimental results in OTs with plasmonic vortices (PVs) using our deep learning network. The specific experimental procedure can be found in the supplemental document. Figs. 7(a-e) show the circumferential rotation of particles near the PVs focal region in the experiment. According to the experimental results, we obtained a set of particle trajectory data through data processing. Fig. 7(f) shows the particle trajectory in this time interval, where A and B represented the starting point and end point of the entire motion process, respectively. The particle trajectory data was inputted into our inverse network to output a set of parameters of optical vortex tweezers from our database. Fig. 7(g) shows the parameters: $(l_1, l_2) = (5, 0)$, $r_0 = 256$ nm, $(x_0, y_0) = (-0.94$ μm, 0 μm). Next, our optical vortex tweezer was established by using these design parameters to achieve particle motion. Finally, we evaluated the performance of our inverse primary network by comparing the particle trajectory based on OV and PV. As shown in Fig. 7(h), a strong match between the targets and simulated designs can be observed, which demonstrates the effectiveness of the network. Based on the above study, on the one hand, we can find a solution in our OV database to satisfy the same particle trajectory achieved based on PV using our network, i.e. PV tweezers on demand This solution is also perspective for particle manipulation in other complex optical fields. On the other hand, as a previous research focus, the characteristics of PVs were difficult to be revealed clearly in earlier studies. Our deep learning network can reveal that PVs have certain characteristics similar to those of OVs, which was shown to be consistent in subsequent studies of PVs using near-field scanning optical microscopy (NSOM) [37].

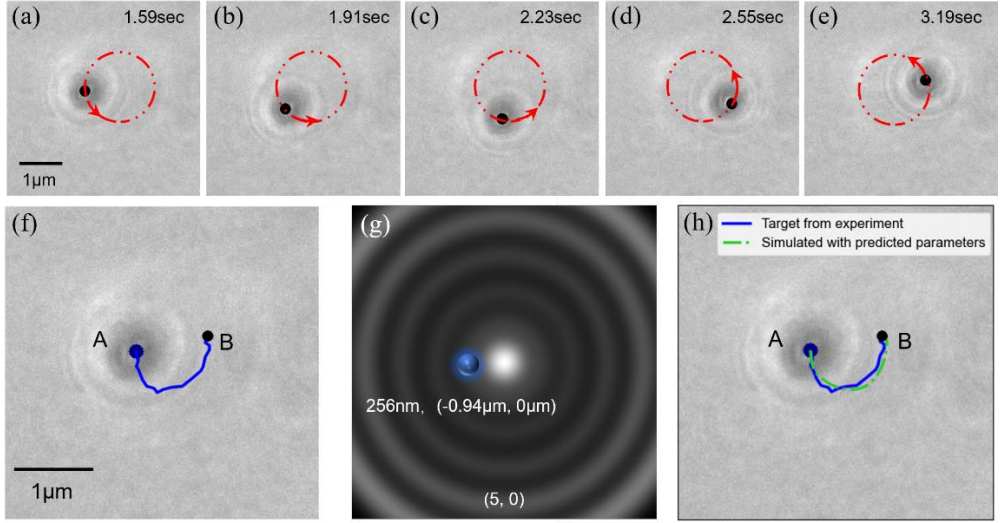

**Fig. 7.** (a-e) Successive frames of a video recording that show the movement of a particle by PV with topological charge $l$=2 (see media). The particle in the red dotted circle is to indicate the clockwise rotation. (f) the detailed particle trajectory obtained from video processing of (a-e) part. (g) The predicted design parameters of optical vortex tweezer by inverse primary network. (h) The comparison results between the particle trajectory of PV and particle trajectory in (g).

## 4. Conclusions

In summary, we proposed a purpose-designed deep-learning framework to analyze and design optical vortex tweezers. The forward subnetwork in our framework can achieve over 98% accuracy while improving computational efficiency by more than 20 times when analyzing optical forces. In further analyzing particle trajectory, the forward primary network in our framework can achieve an accuracy of more than 95.5%. Meanwhile, the inverse primary network in our framework can inverse design optical vortex tweezers on demand, and the accuracy reaches 95.4%. Our framework can be effectively used as a solution for analyzing and designing the optical vortex tweezers. Furthermore, in OTs with OV-like beams, our forward primary network can still predict the motion behaviors of particles with a high accuracy of up to 96.2%. Our framework has a wide applicability, and it is foreseeable that it may be applied in other structured optical tweezers. In addition, our framework provided a solution to analyze and design OTs with PVs, which proves that the optical vortex tweezers design is practicable by pre-defined particle trajectory in practical application using our framework. Thus, our proposed method can provide an effective algorithm platform for the analysis and design of complex OTs environments, it can be widely extended to biophotonics design, optical field analysis, and other fields.

**Funding.** National Natural Science Foundation of China (62275122, 61805119); Natural Science Foundation of Jiangsu Province (BK20180469, BK20180468); Fundamental Research Funds for the Central Universities (30919011275).

**Disclosures.** The authors declare no conflicts of interest.

**Data availability.** Data underlying the results presented in this paper are not publicly available at this time but may be obtained from the authors upon reasonable request.

**Supplemental document.** See Supplement 1 for supporting content.